\documentclass{iopart}
\usepackage{graphicx}
\usepackage{xcolor}
\usepackage{hyperref}

\usepackage[utf8]{inputenc}
\usepackage[T1]{fontenc}
\usepackage{mathptmx}
\usepackage{etoolbox}

\begin{document}

\submitto{\NJP}

\title[Diode Effect for Skyrmions Interacting with Linear Protrusion Defects]{Diode Effect for Skyrmions Interacting with Linear Protrusion Defects
}

\author{J. C. Bellizotti Souza$^{1,*}$,
  C. J. O. Reichhardt$^2$,
  C. Reichhardt$^2$,
  N. P. Vizarim$^1$,
  P. A. Venegas $^3$}

\ead{jc.souza@unesp.br}
\address{$^1$ ``Gleb Wataghin'' Institute of Physics, University of Campinas, 13083-859 Campinas, S\~ao Paulo, Brazil}

\address{$^2$ Theoretical Division and Center for Nonlinear Studies, Los Alamos National Laboratory, Los Alamos, New Mexico 87545, USA}

\address{$^3$ Department of Physics, S\~ao Paulo State University (UNESP), School of Sciences, Bauru 17033-360, SP, Brazil}

\address{$^*$ Corresponding author}

\date{\today}

\begin{abstract}
We simulate collectively interacting skyrmions in a channel with periodic asymmetry, and find a strong diode effect for the skyrmion flow. There is also an asymmetry in the skyrmion annihilation rate for currents applied along the hard or easy substrate asymmetry direction, with a higher annihilation rate for hard direction currents. We map out the diode efficiency as a function of magnetic field and substrate asymmetry angle. We also show that the Magnus force impacts the diode motion and annihilation rate asymmetry by forcing skyrmions into corners of the protrusion geometry.
\end{abstract}


\maketitle

\section{Introduction}
Diode effects are characterized by flow that is easy in
one direction and hindered in the other.
In electric diodes produced with a p-n
junction \cite{Kitai11}, there is
an easy flow of electrons when the junction is positively
polarized, and no flow under negative polarization.
The use of diodes is essential for modern
hardware logic devices.
Diode-like behavior has been observed
in systems ranging from superconductors
\cite{Lyu21, OlsonReichhardt13, Reichhardt10a, Harrington09},
colloidal particles \cite{Reichhardt18c},
fluids \cite{Shou18, Mates14},
photonics \cite{Tocci95, Scalora94, Wang13a},
and thermodynamics \cite{Li04, MartinezPerez15}.
For overdamped systems, such as superconducting vortices,
diode effects can be produced using a combination
of asymmetric potentials, collective
particle interactions, and dc driving
\cite{OlsonReichhardt13, Reichhardt10a},
and the diode effect intensity can be tuned by
varying the number of particles that are present.

More recently, skyrmionic diodes were proposed
for ferromagnetic skyrmion systems
\cite{Feng22, Song21, Jung21, Shu22, Zhao20, Souza22, Souza22a, Souza25b}.
Skyrmions are particle-like magnetic textures that
form triangular lattices in clean samples
\cite{Muhlbauer09, Yu10, Nagaosa13}
and can be set into motion by an
applied spin current \cite{Jonietz10, Iwasaki13a}.
When skyrmions interact with defects or interfaces,
they can experience pinning effects that cause the threshold current
for motion to become finite and nonzero
\cite{Reichhardt22}.
A key feature
that distinguishes skyrmion dynamics from the dynamics of
most other systems is that skyrmion
motion has a strong non-dissipative Magnus force
component \cite{Nagaosa13, Reichhardt22, EverschorSitte18}.
The Magnus force causes skyrmions to
travel at a finite Hall angle with respect
to an external drive and also affects how the skyrmions interact
with interfaces, pinning, and other skyrmions
\cite{Jiang17, Litzius17, Brearton21}.
The presence of the Magnus force
removes the necessity
of collective effects usually required
for diodes in overdamped systems.
The Magnus force also increases the velocity of
skyrmions interacting with defects under external driving
\cite{Souza22, Iwasaki14, Chen17, CastellQueralt19},
making it possible to achieve rapid skyrmion motion
in skyrmionic diode devices.

In recent studies of skyrmion dynamics,
ratchet effects were observed under different
potential landscapes, currents and magnetic fields
\cite{Yamaguchi20, Reichhardt15a, Chen20, Chen19, Reichhardt15b, Ma17, Gobel21, Souza21, Souza24a, Souza25b, Migita20, Moon16}.
Systems in which ratchet effects appear under ac driving
are promising candidates for producing diode
effects when collective interactions are introduced.
For example, after ratchet effects were identified for
individual skyrmions moving over a funnel shaped drive
under ac driving \cite{Souza21},
it was shown that multiple interacting skyrmions in the same
potential could generate diode effects under dc driving.
\cite{Souza22}.
Ratchet effects of both superconducting
vortices \cite{Wells17} and magnetic skyrmions \cite{Souza24a}
interacting with a linear protrusion defect arrangement
have been demonstrated under ac driving.
Here, we show that the same type of
linear protrusion defect array
can give diode motion for multiple interacting skyrmions
under dc driving.

We use atomistic based simulations to
investigate skyrmion dynamics in the presence
of an array of linear protrusion defects under dc driving.
Under positive currents, the skyrmions flow in the
$+x$ direction with velocities of order 20~m~s$^{-1}$,
and the average velocity increases linearly with applied current.
For negative currents the skyrmions can move
along the $-x$ direction at reduced velocities.
This $-x$ motion occurs only for a small current interval, and the
flow disappears for most negative current values.
The magnetic skyrmion
diode effect we observe
is characterized by rapid flow along the easy or $+x$
direction and diminished or vanishing flow along the hard or $-x$
direction.
The rate at which skyrmions annihilate also depends on the
direction of the skyrmion flow, and is less rapid for $+x$ flow
and more rapid for $-x$ flow, with complete skyrmion annihilation
occurring for large negative currents.
The skyrmion trajectories indicate that
interactions with the protrusion defects generate a strong
Magnus velocity boost for easy direction flow, but that during
hard direction flow the moving skyrmions instead interact
with pinned skyrmions.
The angle of the linear protrusions strongly affects
the velocity of the $+x$ flow, since protrusion defects with
small angles are more closely aligned with the drag forces on
the skyrmions and can more effectively enhance the skyrmion
velocity under positive currents. The linear protrusion angle
has little to no effect on the velocity of the $-x$ flow.
Larger protrusion angles reduce the annihilation probability for
both flow directions.
Reducing the magnetic field does not modify the $+x$ flow, but
the resulting softer skyrmions can deform to a greater extent
during $-x$ flow, enhancing the motion. The skyrmion annihilation
rate is also reduced when the magnetic field is lowered,
indicating that
although larger magnetic fields stabilize stiffer skyrmions,
these stiff skyrmions are also brittle and are more susceptible
to annihilation.

\section{Simulation}

\begin{figure}
  \centering
  \includegraphics[width=\columnwidth]{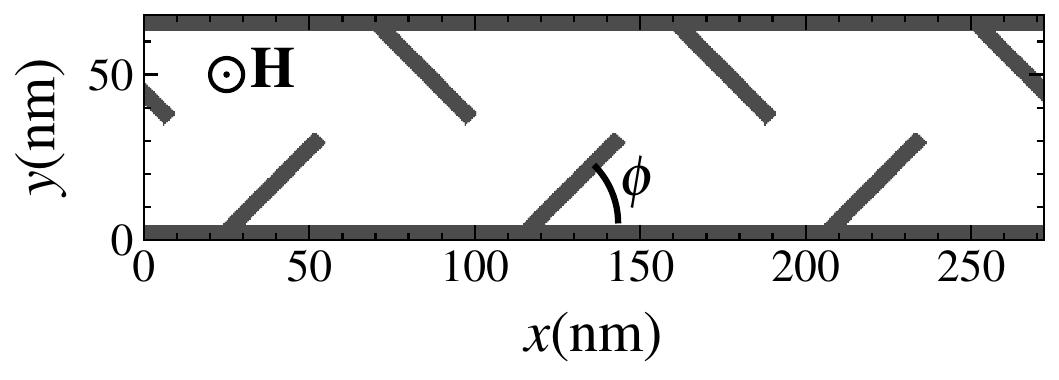}
  \caption{Schematic of the defect arrangement used in this work.
    Gray regions indicate high anisotropy regions. The angle
    $\phi$ is defined as the angle between the $x$ axis and the
    protrusion. The magnetic field points in the positive $z$
    direction, ${\bf H}=H\hat{\bf z}$.}
  \label{fig:1}
\end{figure}

We use atomistic simulations, which can capture the dynamics of
individual atomic magnetic moments \cite{Evans18}, to
model a ferromagnetic ultrathin film capable of holding
N{\'e}el skyrmions. Our sample has dimensions of 272 nm $\times$ 68 nm with
periodic boundary conditions along the $x$ direction. We apply a
magnetic field perpendicular to the sample along the $+z$
direction and work at zero
temperature, $T=0$ K.

The Hamiltonian governing the atomistic dynamics is given by
\cite{Iwasaki13, Evans18, Iwasaki13a}:

\begin{eqnarray}\label{eq:1}
  \mathcal{H}=&-\sum_{\langle i,
    j\rangle}J_{ij}\mathbf{m}_i\cdot\mathbf{m}_j -\sum_{\langle i,
    j\rangle}\mathbf{D}_{ij}\cdot\left(\mathbf{m}_i\times\mathbf{m}_j\right)\\\nonumber
  &-\sum_i\mu\mathbf{H}\cdot\mathbf{m}_i - \sum_{i} K_i\left(\mathbf{m}_i\cdot\hat{\mathbf{z}}\right)^2 \ . \\\nonumber
\end{eqnarray}

The ultrathin film is modeled as a square arrangement of atoms with a
lattice constant $a=0.5$ nm. The first term on the right hand side
of Eq.~(1) is
the exchange interaction with an exchange constant of $J_{ij}=J$
between magnetic moments $i$ and $j$. The second term is the
interfacial Dzyaloshinskii–Moriya interaction, where
$\mathbf{D}_{ij}=D\mathbf{\hat{z}}\times\mathbf{\hat{r}}_{ij}$ is the
Dzyaloshinskii–Moriya vector between magnetic moments $i$ and $j$ and
$\mathbf{\hat{r}}_{ij}$ is the unit distance vector between sites $i$
and $j$. Here, $\langle i, j\rangle$ indicates that the sum is over
only the first neighbors of the $i$th magnetic moment. The third term
is the Zeeman interaction with an applied external magnetic field
$\mathbf{H}$. Here $\mu=g\mu_B$ is the magnitude of the magnetic
moment, $g=|g_e|=2.002$ is the electron $g$-factor, and
$\mu_B=9.27\times10^{-24}$~J~T$^{-1}$ is the Bohr magneton. The last
term represents the sample perpendicular magnetic anisotropy (PMA),
where we use two PMA constants: $K=0.02J$ for $i\notin P$, and $K=5J$
for $i\in P$, with $P$ being the set of defects composing the
linear protrusion arrangement shown in Fig.~\ref{fig:1}.
In ultrathin films, long-range dipolar interactions
act as a PMA (see Supplemental of Wang {\it et al.} \cite{Wang18}),
and therefore merely effectively shift the PMA values.

The time evolution of atomic magnetic moments is obtained using the
Landau-Lifshitz-Gilbert (LLG)
equation \cite{Seki16, Gilbert04}:

\begin{equation}\label{eq:2}
  \frac{\partial\mathbf{m}_i}{\partial
    t}=-\gamma\mathbf{m}_i\times\mathbf{H}^\mathrm{eff}_i
  +\alpha\mathbf{m}_i\times\frac{\partial\mathbf{m}_i}{\partial t}
  +\frac{pa^3}{2e}\left(\mathbf{j}\cdot\nabla\right)\mathbf{m}_i \ .
\end{equation}
Here $\gamma=1.76\times10^{11}~$T$^{-1}$~s$^{-1}$ is the electron
gyromagnetic ratio,
$\mathbf{H}^\mathrm{eff}_i=-\frac{1}{\mu}\frac{\partial \mathcal{H}}{\partial \mathbf{m}_i}$
is the effective magnetic field including all interactions from the Hamiltonian, $\alpha$ is the
phenomenological damping introduced by Gilbert, and the last term is
the adiabatic spin-transfer-torque (STT) caused by application of an in
plane spin polarized current, where $p$ is the spin polarization, $e$
the electron charge, and $\mathbf{j}=j\hat{\mathbf{y}}$ the applied
current density. Use of this STT expression implies that the
conduction electron spins are always parallel to the magnetic moments
$\mathbf{m}$ \cite{Iwasaki13,Zang11}.
In this work we only consider adiabatic STT contributions, which
corresponds to dragging forces perpendicular to ${\bf j}$ \cite{Iwasaki13a, Feilhauer20},
so dragging forces are along $\hat{\bf x}$ in our system.
It is possible to include non-adiabatic STT contributions in Eq.~\ref{eq:2};
however, non-adiabatic STT does not appreciably affect
the dynamics of nanoscale skyrmions at small driving forces
\cite{Litzius17}, which is the regime we consider.
Other types of torques, such as spin orbit torques, can also be investigated,
but are not explored in the present work.

The skyrmion velocity is computed using the emergent electromagnetic
fields \cite{Seki16, Schulz12}:

\begin{eqnarray}\label{eq:5}
  E^\mathrm{em}_i=\frac{\hbar}{e}\mathbf{m}\cdot\left(\frac{\partial
    \mathbf{m}}{\partial i}\times\frac{\partial \mathbf{m}}{\partial
    t}\right)\\ B^\mathrm{em}_i=\frac{\hbar}{2e}\varepsilon_{ijk}\mathbf{m}\cdot\left(\frac{\partial
    \mathbf{m}}{\partial j}\times\frac{\partial \mathbf{m}}{\partial
    k}\right) \ ,
\end{eqnarray}
where $\varepsilon_{ijk}$ is the totally anti-symmetric tensor. The
skyrmion drift velocity, $\mathbf{v}_d$, is then calculated using
$\mathbf{E}^\mathrm{em}=-\mathbf{v}_d\times\mathbf{B}^\mathrm{em}$.

We fix the following values in our simulations:
$\alpha=0.3$ and $p=-1.0$. The material parameters are
$J=1$ meV and $D=0.2J$.
The numerical integration of Eq.~\ref{eq:2} is
performed using a fourth order Runge-Kutta method.
For each set of $\mu H$ and $\phi$, the system is initialized
with a triangular skyrmion lattice,
and then relaxed for 10 ns to allow the skyrmions to adjust
to the presence of the linear
protrusion defects shown in Fig.~\ref{fig:1}.
To ensure a steady state for measurement we evolve
Eq.~\ref{eq:2} over 200 ns.

\section{Diode effect}

\begin{figure}
  \centering
  \includegraphics[width=\columnwidth]{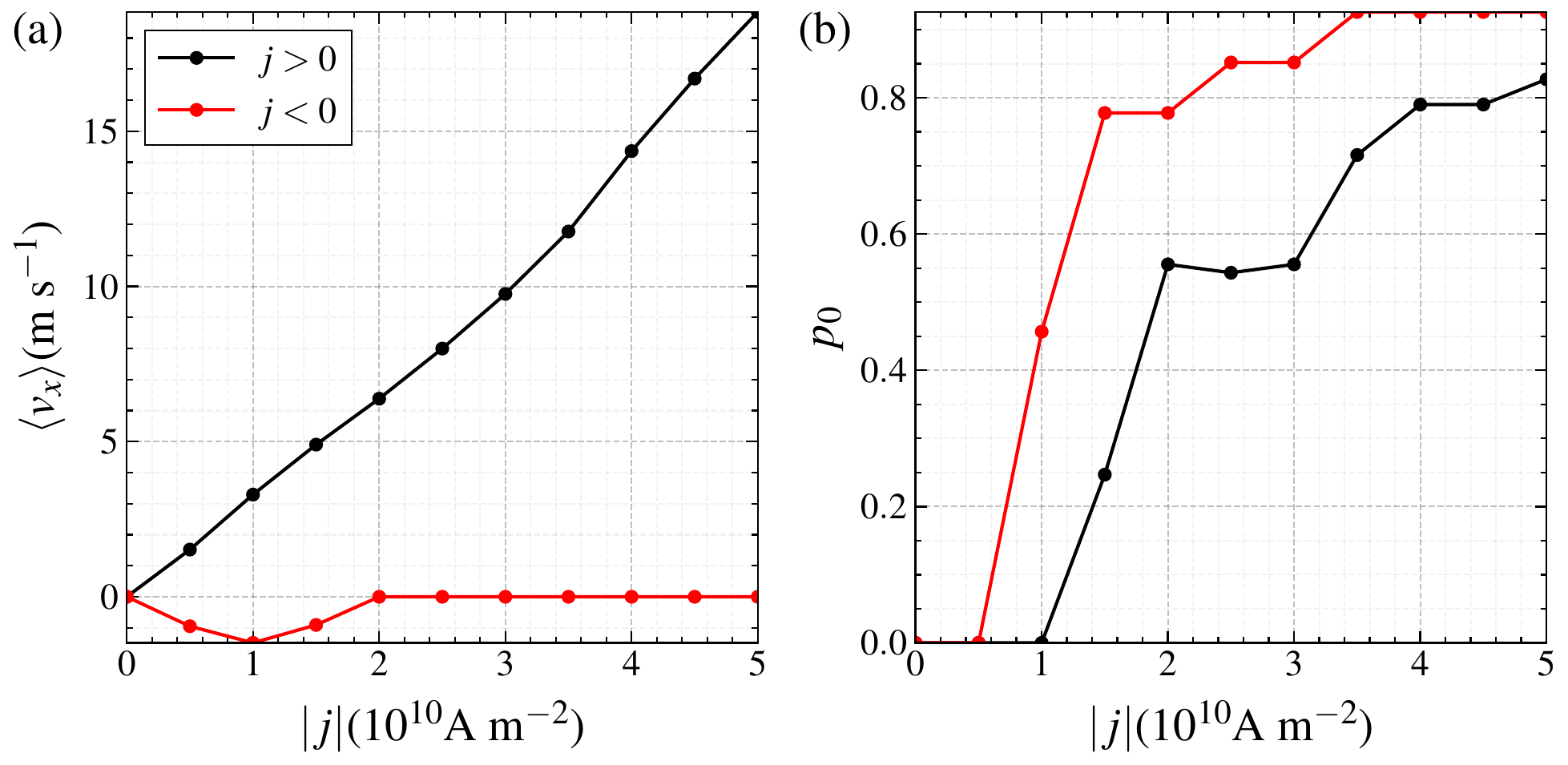}
  \caption{(a) Skyrmion average velocity along $x$, $\langle v_x\rangle$,
    and (b) annihilation probability $p_0$ vs
    the magnitude $j$ of the applied current
    in a sample with $\phi=45^\circ$
    and $\mu H=0.5D^2/J$.
    Black: positive currents $j>0$; red: negative currents $j<0$.}
  \label{fig:2}
\end{figure}

\begin{figure}
  \centering
  \includegraphics[width=\columnwidth]{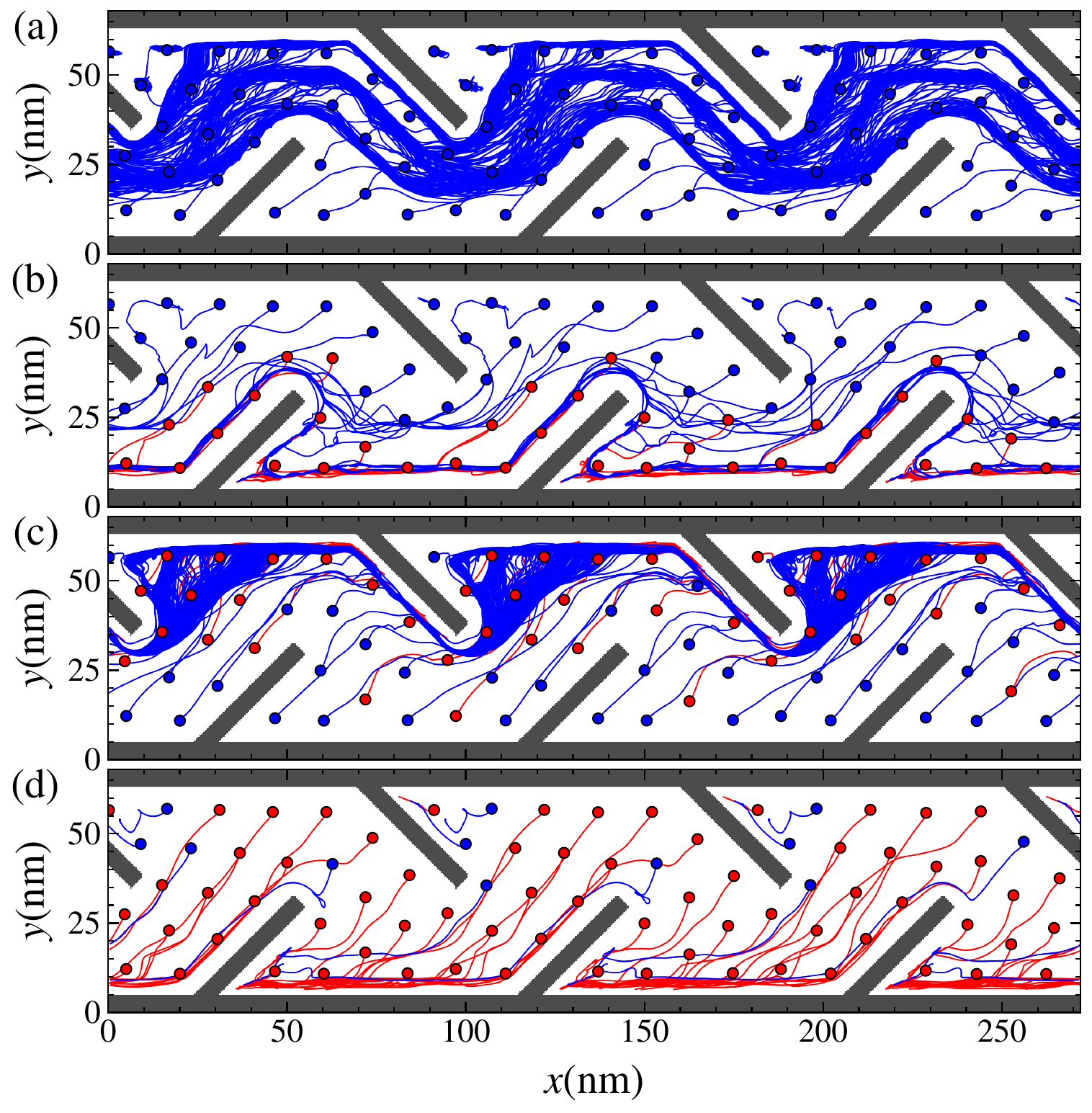}
  \caption{Skyrmion trajectories from the system
    in Fig.~\ref{fig:2} with $\phi=45^\circ$ and $\mu H=0.5D^2/J$ at
    (a) $j=1\times10^{10}$~A~m$^{-2}$,
    (b) $j=-1\times10^{10}$~A~m$^{-2}$,
    (c) $j=2.5\times10^{10}$~A~m$^{-2}$, and
    (d) $j=-2.5\times10^{10}$~A~m$^{-2}$.
    Blue dots and trajectories: skyrmions that did not annihilate.
    Red dots and trajectories: skyrmions that annihilated during
    the flow.
    Animations of the skyrmion dynamics can be found
    in the supplemental material \cite{suppl}.
  }
  \label{fig:3}
\end{figure}

We first consider a diode effect in
a sample with a magnetic field of $\mu H=0.5D^2/J$
and a protrusion angle of $\phi=45^\circ$.
In Fig.~\ref{fig:2}(a) we plot
the average skyrmion velocity $\langle v_x\rangle$ versus current amplitude $j$
for positive and negative currents.
Under a positive current with $j>0$, $\langle v_x\rangle$ increases
linearly with $j$,
reaching $\langle v_x\rangle\approx 19$~m~s$^{-1}$
at $j=5\times10^{10}$~A~m$^{-2}$.
For negative currents with $j<0$,
$|\langle v_x\rangle|$ slowly increases until
$j=-1\times10^{10}$~A~m$^{-2}$, and then it decreases to the value
$|\langle v_x\rangle|=0$~m~s$^{-1}$ at and above $|j|=2\times10^{10}$~A~m$^{-2}$.
The significant difference in absolute velocity between
positive and negative currents indicates that
a diode effect is present, where the flow
is easier along the $+x$ direction and harder along the $-x$ direction.

The plot of the annihilation probability $p_0$ versus current amplitude
$j$ in Fig.~\ref{fig:2}(b) again indicates that positive and negative
currents produce differing behavior.
Under a positive current,
we find nonzero skyrmion annihilation
above $j=1\times10^{10}$~A~m$^{-2}$, while the onset of
annihilation begins at a lower current amplitude of
$|j|=0.5\times10^{10}$~A~m$^{-2}$ for negative currents.
In each case, $p_0$ increases rapidly at lower $|j|$ and then
approaches a saturation value of
$p_0=1$ with increasing $|j|$.
For negative currents, the saturation of $p_0$ is complete at and
above $|j_s|=3.5\times10^{10}$~A~m$^{-2}$, but we do not reach
complete saturation of $p_0$ in the range of positive currents
considered here.
For all values of $|j|$,
the rate of skyrmion annihilation is
greater for negative than for positive currents.

To better describe the annihilation process,
in Fig.~\ref{fig:3} we plot the skyrmion
trajectories at selected values of $j$.
There is no skyrmion annihilation at $j=1\times10^{10}$~A~m$^{-2}$ in
Fig.~\ref{fig:3}(a),
and the skyrmions interact strongly
with the linear protrusion defects while flowing, resulting in a
velocity increase via
the Magnus velocity boost effect
\cite{Souza22, Iwasaki14, Chen17, CastellQueralt19}.
Three well defined flow channels are present, with the first
and second closely following the protrusion defect profile, and the
third interacting with skyrmions in stagnant areas of the channel.
Near the protrusions,
we observe compression of the skyrmions similar to those reported
by Souza {\it et al.} \cite{Souza23a}, where a skyrmion size
gradient is present.
In Fig.~\ref{fig:3}(b) at
$j=-1\times10^{10}$~A~m$^{-2}$, a portion of the
skyrmions are annihilated during the motion.
The skyrmions flow primarily in a single channel along the
bottom of the sample.
A steady state flow is established in which one skyrmion becomes trapped
in the acute corner of the bottom linear protrusion defects. This skyrmion
undergoes large periodic fluctuations in size as the other skyrmions in
the moving channel travel past. At the same time, a trio of skyrmions
becomes pinned in each acute corner of the upper linear protrusion
defects as a consequence of the chirality exhibited by the skyrmions when
interacting with defects \cite{Zhang23a}.
At $j=2.5\times10^{10}$~A~m$^{-2}$ in
Fig.~\ref{fig:3}(c),
an even greater amount of skyrmion annihilation
occurs and the formation of more than one flowing lane becomes
impossible. Due to the sign of the current, this lane follows the
contour of the upper linear protrusion defects.
When the current is reversed to
$j=-2.5\times10^{10}$~A~m$^{-2}$,
Fig.~\ref{fig:3}(d) shows that nearly all of the skyrmions annihilate,
and the surviving skyrmions become pinned in the upper and lower acute
angles of the linear protrusion defects.

\section{Varied protrusion angle $\phi$}

\begin{figure}
  \centering
  \includegraphics[width=\columnwidth]{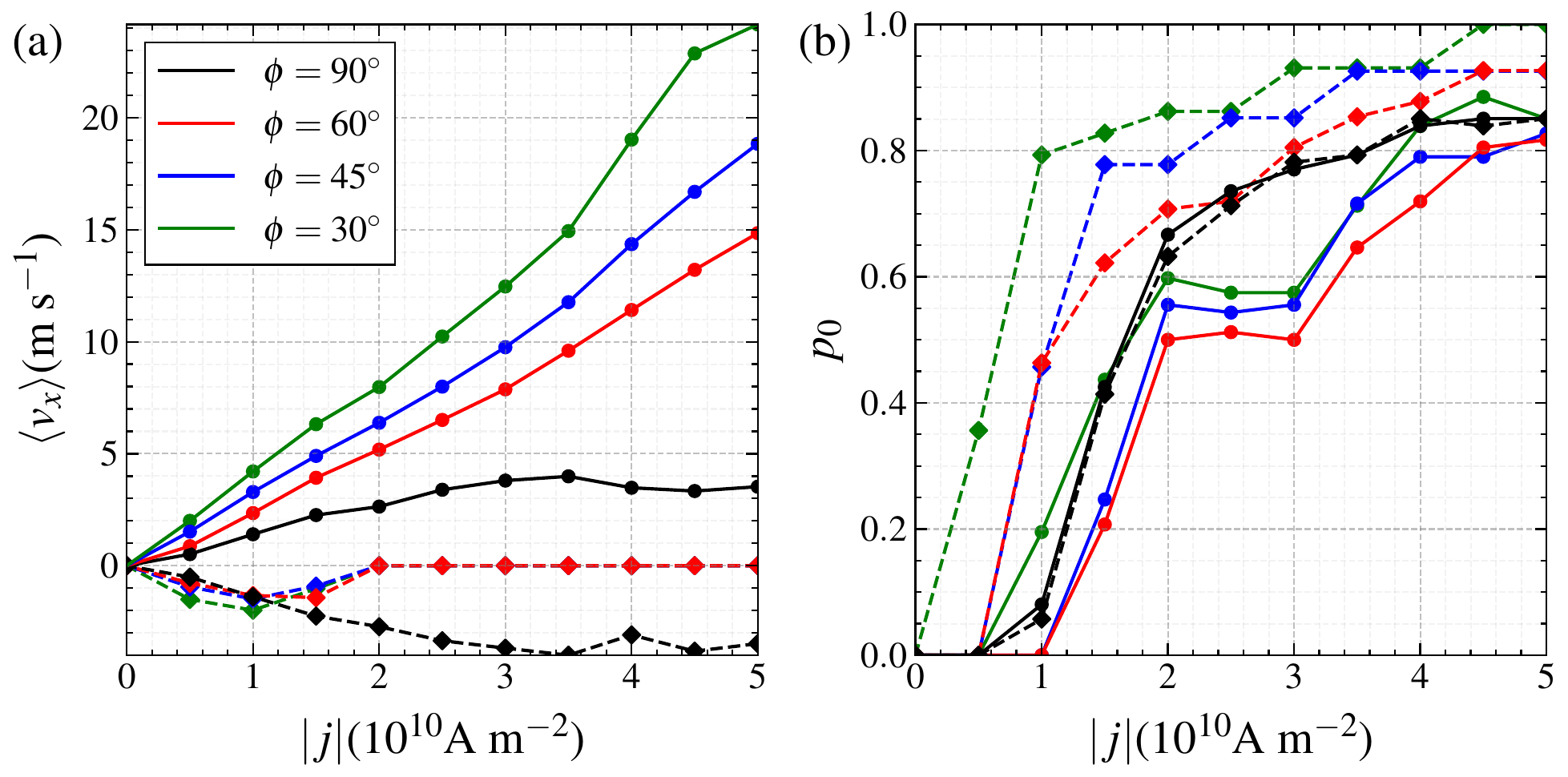}
  \caption{(a) $\langle v_x\rangle$ and (b)
    $p_0$ vs the magnitude of the applied current $j$ in a sample with
    $\mu H=0.5D^2/J$ at different
    protrusion angles
    $\phi=90^\circ$ (black),
    $\phi=60^\circ$ (red),
    $\phi=45^\circ$ (blue), and
    $\phi=30^\circ$ (green).
    Solid lines with circles represent
    positive currents with $j>0$, while
    dashed lines with diamonds represent
    negative currents with $j<0$.
  }
  \label{fig:4}
\end{figure}

\begin{figure}
  \centering
  \includegraphics[width=0.5\columnwidth]{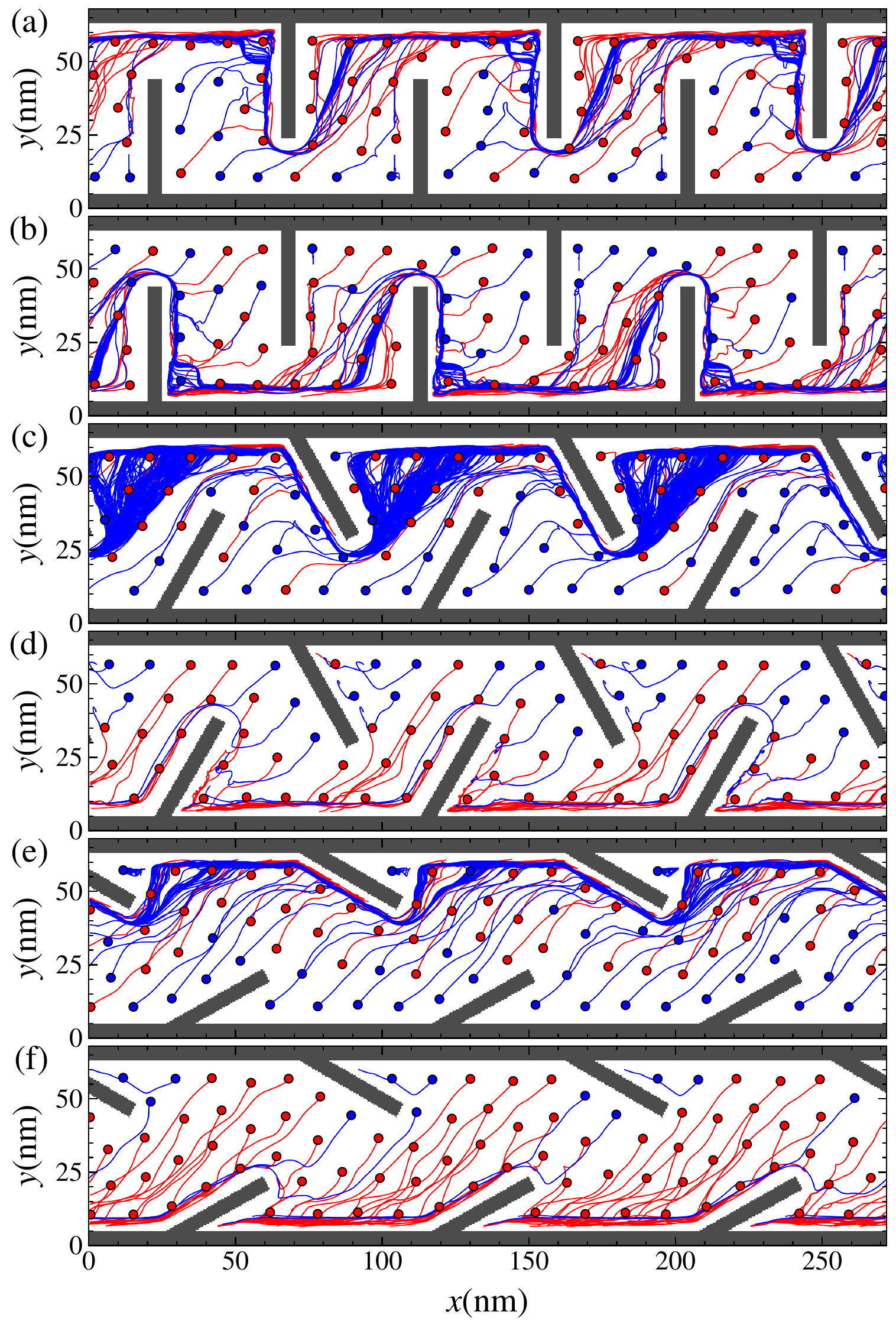}
  \caption{Skyrmion trajectories from the system
    in Fig.~\ref{fig:4} with $\mu H=0.5D^2/J$ at
    (a) $j=2.5\times10^{10}$~A~m$^{-2}$ and $\phi=90^\circ$,
    (b) $j=-2.5\times10^{10}$~A~m$^{-2}$ and $\phi=90^\circ$,
    (c) $j=2.5\times10^{10}$~A~m$^{-2}$ and $\phi=60^\circ$,
    (d) $j=-2.5\times10^{10}$~A~m$^{-2}$ and $\phi=60^\circ$,
    (e) $j=2.5\times10^{10}$~A~m$^{-2}$ and $\phi=30^\circ$, and
    (f) $j=-2.5\times10^{10}$~A~m$^{-2}$ and $\phi=30^\circ$.
    Blue dots and trajectories: skyrmions that did not annihilate.
    Red dots and trajectories:
    skyrmions that annihilated during the flow.
    Animations of the skyrmion dynamics can be found in the
    supplemental material \cite{suppl}.
  }
  \label{fig:5}
\end{figure}

We next study the effect on
$\langle v_x\rangle$ and $p_0$ of modifying the
protrusion angle $\phi$.
Figure~\ref{fig:4}(a) shows
$\langle v_x\rangle$ versus $j$
for different values of $\phi$.
Under positive currents $j>0$,
the magnitude of the velocity decreases with increasing $\phi$.
For example, at $j=5\times10^{10}$~A~m$^{-2}$, we find
$\langle v_x\rangle\approx 24$~m~s$^{-1}$ when $\phi=30^\circ$,
$\langle v_x\rangle\approx 19$~m~s$^{-1}$ when $\phi=45^\circ$,
$\langle v_x\rangle\approx 15$~m~s$^{-1}$ when $\phi=60^\circ$,
and $\langle v_x\rangle\approx 3.5$~m~s$^{-1}$ when $\phi=90^\circ$.
In contrast, for negative currents
$j<0$, the velocity is only weakly sensitive
to the value of $\phi$, and
in almost all cases
$\langle v_x\rangle=0$~m~s$^{-1}$ when $|j|\geq 2\times10^{10}$~A~m$^{-2}$.
The exception is that at
$\phi=90^\circ$, the $+j$ and $-j$ behaviors
are almost perfectly mirrored across the zero velocity line. The small
differences that we find in these two curves are expected to appear
as a consequence
of the skyrmion chirality when interacting with defects \cite{Zhang23a}.

In Fig.~\ref{fig:4}(b) we plot the annihilation
probability $p_0$ versus $j$ for different
protrusion angles $\phi$.
The positive $j>0$ current behavior for all $\phi$ values
is very similar to the behavior of the $\phi=45^\circ$ system
in Fig.~\ref{fig:2}(b), where $p_0$ increases rapidly with increasing
$j$ at low currents before beginning to saturate toward the value
$p_0=1$ at larger $j$.
For a given value of $j$,
the annihilation probability decreases as $\phi$ increases.
The onset of skyrmion annihilation falls at
$j = 0.5\times10^{10}$~A~m$^{-2}$
for $\phi=30^\circ$ and $90^\circ$, but
shifts to higher values of
$j = 1\times10^{10}$~A~m$^{-2}$.
for $\phi=45^\circ$ and $60^\circ$.
Figure~\ref{fig:2}(b) shows that
negative currents with $j<0$ also give behavior
similar to that shown in Fig.~\ref{fig:2}(b), with
a more rapid increase in $p_0$ with increasing $|j|$ at low $|j|$
compared to the positive current system,
followed by a saturation to $p_0=1$ at large $|j|$.
We again observe that, for a fixed value of $j$, increasing $\phi$
decreases the annihilation probability.
Skyrmion annihilation occurs for all negative current values
when $\phi=30^\circ$,
while at $\phi=45^\circ$, $60^\circ$, and $90^\circ$,
the onset of annihilation is at $|j|=0.5\times10^{10}$~A~m$^{-2}$.
For all simulated values of $\phi$,
the annihilation rate is greater for negative currents than for
positive currents, with the exception of
$\phi=90^\circ$,
where the annihilation for $j>0$ and $j<0$
is approximately the same over the entire range of $j$ considered.

In order to better understand the results
of Fig.~\ref{fig:4},
in Fig.~\ref{fig:5}
we plot representative skyrmion trajectories for
different $\phi$ values at a fixed current magnitude of
$|j|=2.5\times10^{10}$~A~m$^{-2}$.
In Fig.~\ref{fig:5}(a), at
$\phi=90^\circ$ and a positive $j>0$ current,
the majority of the skyrmions annihilate, while
the remaining skyrmions form a single flowing channel following
the upper contour of the linear protrusion defects.
Some skyrmions remain
pinned along the lower region of the sample
as a consequence
of their interactions with the defects.
For a negative $j<0$ current at $\phi=90^\circ$, shown in
Fig.~\ref{fig:5}(b), a similar behavior occurs with the roles of
the top and bottom walls of the sample reversed.
For both $\phi=90^\circ$ systems, when skyrmions
interact with the corners at which the
linear protrusions meet the confining upper and lower walls,
their velocities are increased as a result of a Magnus
velocity boost. The boost comes only from the corners and
does not occur along the length of the linear protrusions.
In Fig.~\ref{fig:5}(c), for a positive $j>0$ current in a
system with $\phi=60^\circ$,
approximately half of the skyrmions annihilate.
The surviving skyrmions gradually assemble into a single flowing
channel following the upper contour of the protrusion defects.
Here the Magnus velocity boost becomes very important, since it
is produced not only at the corners where the linear protrusions
emerge from the boundary walls, but also along the length of the
linear protrusion defects.
For negative $j<0$ currents at $\phi=60^\circ$,
Fig.~\ref{fig:5}(d) shows that
the majority of the skyrmions annihilate, while the surviving
skyrmions become pinned inside the upper contour of the protrusions
and cease to move.
At $\phi=30^\circ$ and positive $j>0$ currents in
Fig.~\ref{fig:5}(e),
approximately 60\% of the skyrmions
annihilate.
We again find that the surviving skyrmions travel along the upper
contour of the protrusion defects, similar to the behavior in
Figs.~\ref{fig:3}(a, c) and Fig.~\ref{fig:5}(c).
The skyrmion velocity is enhanced by the Magnus velocity
boost, but the skyrmions have a higher velocity than in the
$\phi=45^\circ$ and $\phi=60^\circ$ samples,
indicating that the magnitude of the
Magnus velocity boost is $\phi$ dependent.
In Fig.~\ref{fig:5}(f) at $\phi=30^\circ$ and negative
$j<0$ current, a majority of the skyrmions annihilate and
the remaining skyrmions become pinned by the protrusion defects.

\section{Varied Magnetic field $\mu H$}

\begin{figure}
  \centering
  \includegraphics[width=\columnwidth]{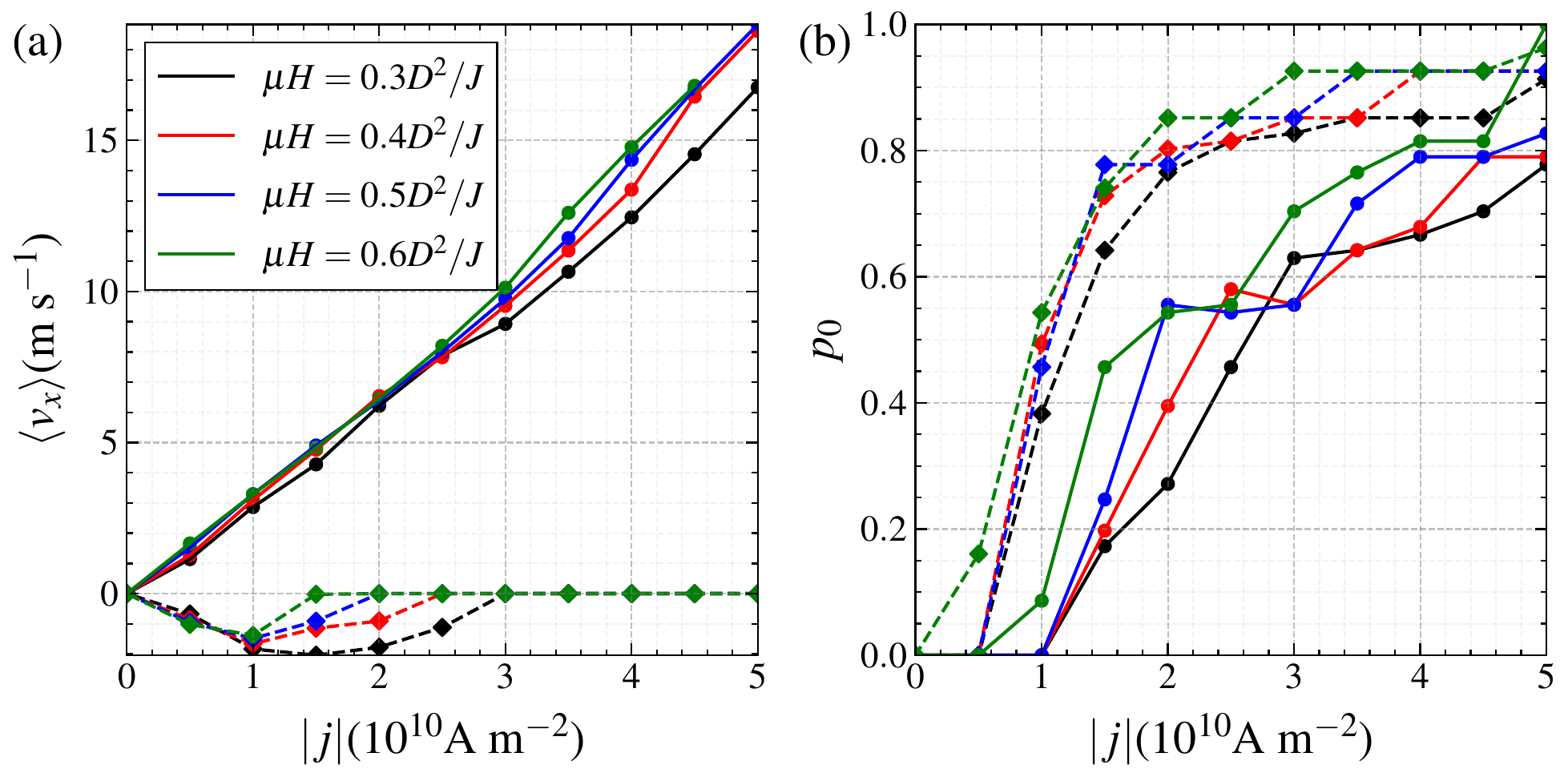}
  \caption{(a) $\langle v_x\rangle$ and (b)
    $p_0$ vs the magnitude of the applied current $j$ in a sample with
    $\phi=45^\circ$ at different
    magnetic fields,
    $\mu H=0.3D^2/J$ (black),
    $\mu H=0.4D^2/J$ (red),
    $\mu H=0.5D^2/J$ (blue), and
    $\mu H=0.6D^2/J$ (green).
    Solid lines with circles represent
    positive currents with $j>0$, while
    dashed lines with diamonds represent
    negative currents with $j<0$.
  }
  \label{fig:6}
\end{figure}

\begin{figure}
  \centering
  \includegraphics[width=0.5\columnwidth]{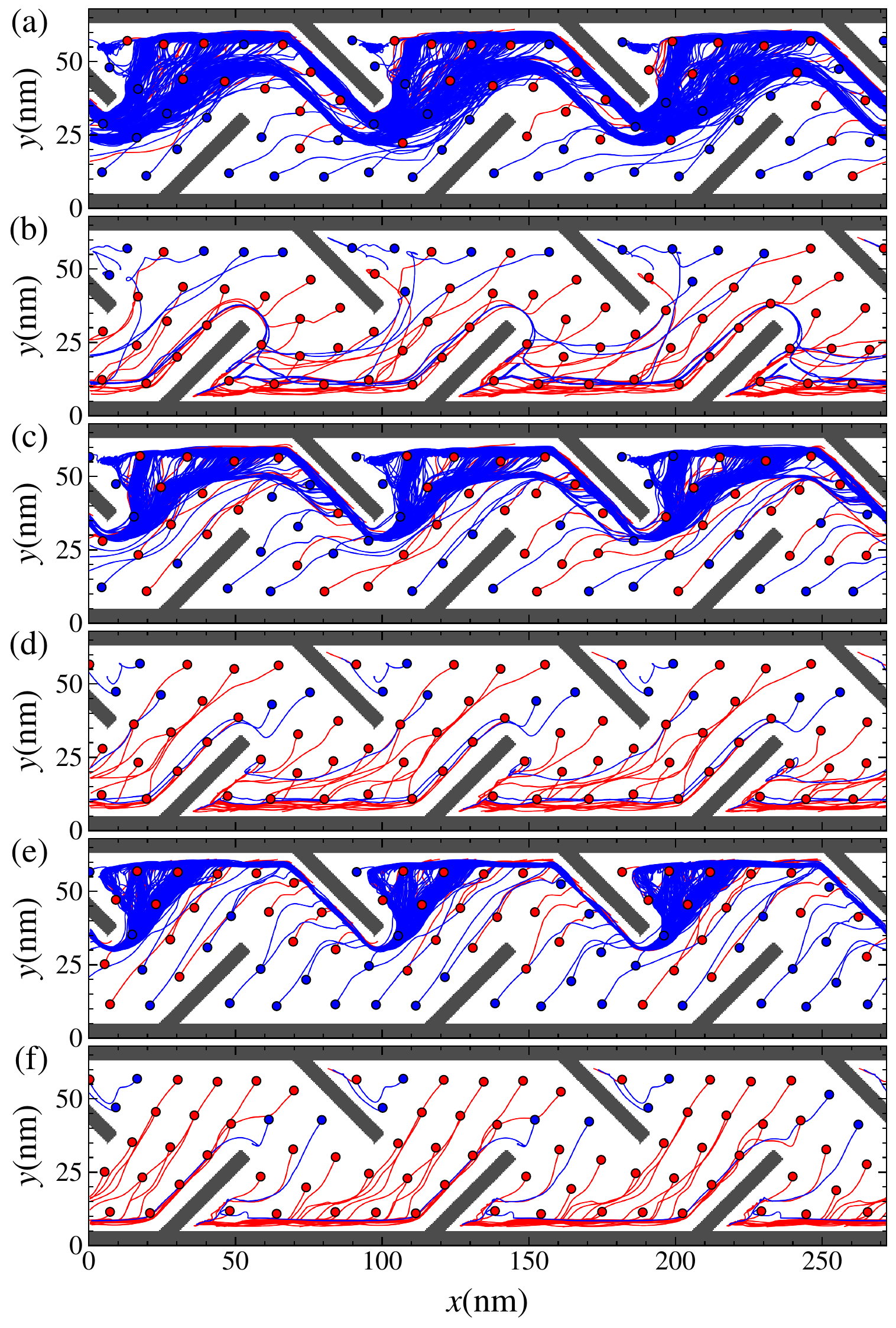}
  \caption{Skyrmion trajectories from the system
    in Fig.~\ref{fig:6} with $\phi=45^\circ$ at
    (a) $j=2.5\times10^{10}$~A~m$^{-2}$ and $\mu H=0.3D^2/J$,
    (b) $j=-2.5\times10^{10}$~A~m$^{-2}$ and $\mu H=0.3D^2/J$,
    (c) $j=2.5\times10^{10}$~A~m$^{-2}$ and $\mu H=0.4D^2/J$,
    (d) $j=-2.5\times10^{10}$~A~m$^{-2}$ and $\mu H=0.4D^2/J$,
    (e) $j=2.5\times10^{10}$~A~m$^{-2}$ and $\mu H=0.6D^2/J$,
    and
    (f) $j=-2.5\times10^{10}$~A~m$^{-2}$ and $\mu H=0.6D^2/J$.
    Blue dots and trajectories: skyrmions that did not annihilate.
    Red dots and trajectories: skyrmions that annihilated during
    the flow.
    Animations of the skyrmion dynamics can be found in the
    supplemental material \cite{suppl}.
  }
  \label{fig:7}
\end{figure}

Next we fix $\phi=45^\circ$ and vary $\mu H$ for
different values of $j$.
Figure~\ref{fig:6}(a) shows $\langle v_x\rangle$ versus
$j$ at applied fields of $\mu H=0.3D^2/J$, $0.4D^2/J$, $0.5D^2/J$,
and $0.6D^2/J$.
Under positive currents $j>0$,
$\langle v_x\rangle$ is affected very little by changes to the
magnetic field,
unlike the sensitivity
of $\langle v_x\rangle$ to
the protrusion angle $\phi$ displayed in Fig.~\ref{fig:4}(a).
A small amount of variation in $\langle v_x\rangle$ appears
only at large values of $j$, where increasing $\mu H$ modestly
increases the velocity.
Much more significant changes in $\langle v_x\rangle$ appear
under negative currents $j<0$, where
both the maximum value of
$|\langle v_x\rangle|$ and the range of currents over which the
velocity is nonzero increase with decreasing magnetic
field.
There is a small increase in the maximum velocity from
$|\langle v_x\rangle|\approx0.3$~m~s$^{-1}$ at $\mu H \geq 0.4D^2/J$
to
$|\langle v_x\rangle|\approx0.4$~m~s$^{-1}$ at $\mu H = 0.3D^2/J$.
The window of nonzero velocity expands significantly as
the magnetic field decreases.
For $\mu H=0.3D^2/J$, $0.4D^2/J$, $0.5D^2/J$, and $0.6D^2/J$,
the width $\Delta j$ of the finite velocity window is,
respectively,
$\Delta j=3\times10^{10}~\mathrm{A~m}^{-2}$,
$2.5\times10^{10}~\mathrm{A~m}^{-2}$,
$2\times10^{10}~\mathrm{A~m}^{-2}$,
and $1.5\times10^{10}~\mathrm{A~m}^{-2}$.
When the magnetic field is smaller, or if the anisotropy is
decreased, the skyrmions become softer
\cite{Souza23a, Liu23}, permitting
deformations to become important for
enabling the flow of skyrmions along the hard or negative $x$
direction.
Our results show that it is possible to tune the magnetic field
for the purpose of
controlling the motion of the skyrmions along the hard
direction, so that the skyrmions are mobile under low fields
but become pinned
under higher fields.

Figure~\ref{fig:6}(b) shows the annihilation
probability $p_0$ versus the magnitude of the applied current $j$
for different values of $\mu H$.
Positive currents with $j>0$ exhibit the same
behavior described in Fig.~\ref{fig:2}(b) and Fig.~\ref{fig:4}(b),
with a rapid increase of $p_0$
with increasing current for low $j$, followed by saturation
towards $p_0=1$ at large $j$.
As the magnetic field increases, the overall magnitude of $p_0$ increases
somewhat, indicating that more annihilation is occurring at larger $\mu H$.
The onset of skyrmion annihilation is weakly affected by changing
the magnetic field,
falling at
$j=0.5\times10^{10}$~A~m$^{-2}$ for
$\mu H=0.6D^2/J$
and at
$j>1\times10^{10}$~A~m$^{-2}$ for
$\mu H<0.6D^2/J$.
For negative currents with $j<0$,
we observe a rapid increase in $p_0$ at
lower $|j|$, and saturation towards $p_0=1$ at
larger $|j|$, similar to the behavior in
Fig.~\ref{fig:2}(b) and Fig.~\ref{fig:4}(b).
We find that, for a given current amplitude,
the annihilation probability for $j<0$
is larger than that for $j>0$.

In Fig.~\ref{fig:7} we plot the skyrmion trajectories
for $|j|=2.5\times10^{10}$~A~m$^{-2}$ and $\phi=45^\circ$
under different magnetic fields $\mu H$.
At
$\mu H=0.3D^2/J$ and $j>0$ in
Fig.~\ref{fig:7}(a),
approximately 45\% of the skyrmions
annihilate.
The remaining
skyrmions exhibit a Magnus velocity boost and
form two channels of flow while interacting with the
defects, with the first
channel following the upper contour
of the protrusion defects, and the second channel following the
contour of the first channel.
The two channels merge and mix each time the skyrmions reach the upper
edge of a protrusion defect, and then reform when the skyrmions arrive
at the next protrusion defect.
For a negative $j<0$ current at
$\mu H=0.3D^2/J$,
Figure~\ref{fig:7}(b) shows
that approximately 80\% of the skyrmions annihilate.
One portion of the remaining skyrmions
becomes pinned near the corners of the protrusion defects, and facilitates
the flow of the rest of the skyrmions
along the hard axis by pushing the moving skyrmions over the ends of
the lower protrusion defects.
Both the pinned and moving skyrmions deform strongly during this motion,
and in some cases the moving skyrmions adapt to the shape of the
protrusion barrier in the course of jumping over the barrier with assistance
from the pinned skyrmions.
In Fig.~\ref{fig:7}(c), for a positive $j>0$ current at $\mu H=0.4D^2/J$,
approximately 60\% of the skyrmions annihilate.
The motion is similar to that found in
Fig.~\ref{fig:7}(a), with the skyrmions experiencing
a Magnus velocity boost while flowing along the easy axis.
There are again two channels
of flow, but now the two channels
merge before the interaction between
the skyrmions and protrusions
ceases,
indicating that the existence of two [or three, Fig.~\ref{fig:3}(a)]
channels of flow depends on the
rigidity of the skyrmions.
Figure~\ref{fig:7}(d) shows a system with $\mu H=0.4D^2/J$
for a negative $j<0$ current, where approximately 80\% of
the skyrmions annihilate.
No persistent motion
appears and all of the surviving skyrmions become pinned.
The larger value of the magnetic field causes the skyrmions
to become more rigid, with
reduced deformations,
which destroys the hard direction skyrmion
flow that was observed at $\mu H=0.3D^2/J$ in Fig.~\ref{fig:7}(b).
In Fig.~\ref{fig:7}(e) at $\mu H=0.6D^2/J$ under a positive $j>0$
current, approximately 55\% of the skyrmions annihilate.
As in Fig.~\ref{fig:7}(a, c), the skyrmions flow along
the easy direction and experience a Magnus velocity boost.
The flow now follows a single channel, with skyrmions
moving closer to the upper contour of the protrusion
defects and walls.
The absence of
a second channel of flow enhances the interactions of
the skyrmions with the protrusions, resulting in the appearance of
a stronger Magnus velocity boost.
Under a negative $j<0$ current at
$\mu H=0.6D^2/J$,
Fig.~\ref{fig:7}(f) shows that
approximately 85\% of
the skyrmions annihilate. As was the case in
Fig.~\ref{fig:7}(d), no persistent motion occurs due to the
increased rigidity of the skyrmions at the larger magnetic field.

\begin{figure}
  \centering
  \includegraphics[width=\columnwidth]{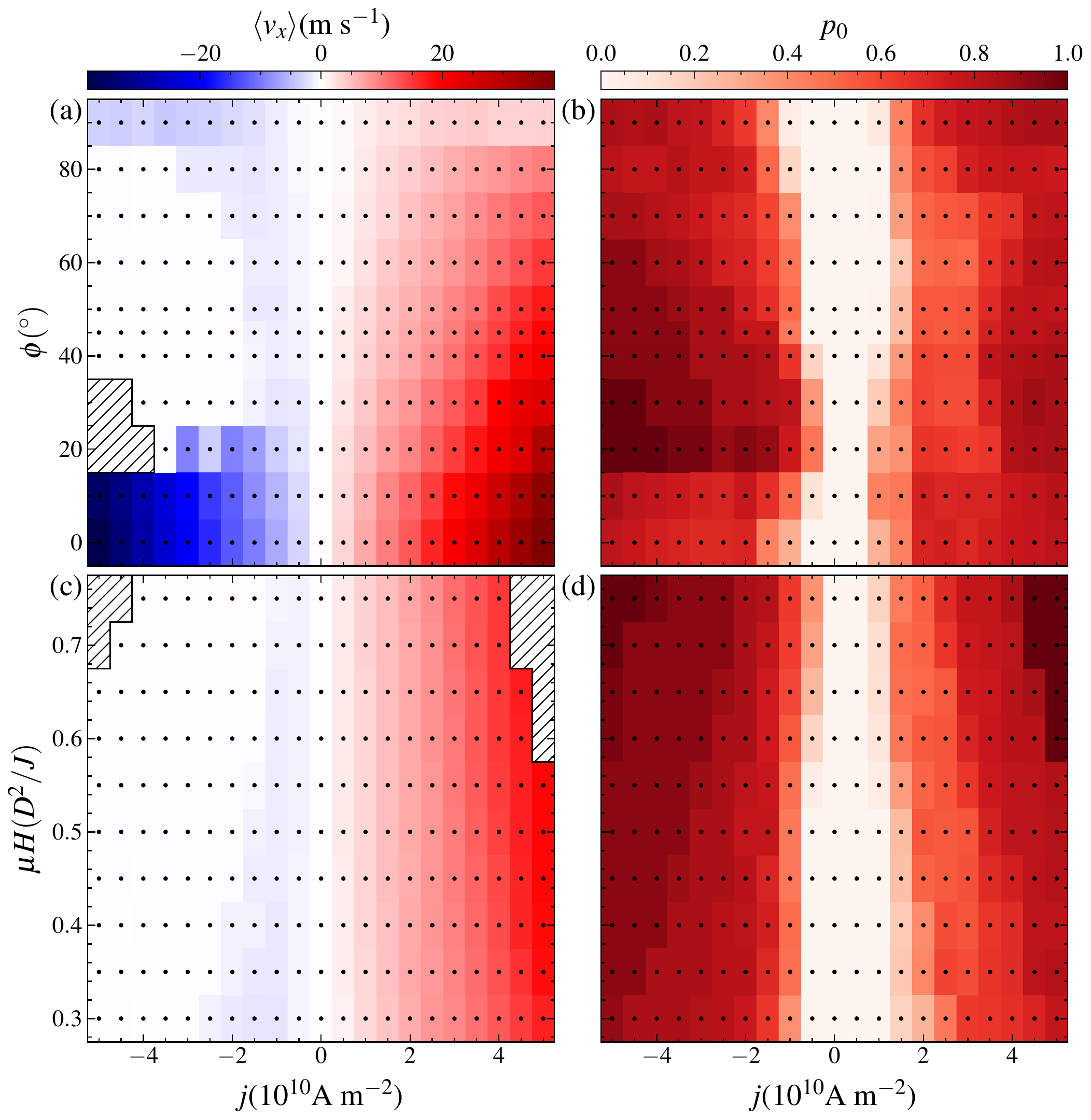}
  \caption{Heat maps of (a, c) the skyrmion
    average velocity along $x$, $\langle v_x\rangle$
    and (b, d) the annihilation probability $p_0$
    plotted as a function of
    (a, b) protrusion angle $\phi$ vs applied current $j$
    or (c, d) magnetic field $\mu H$ vs applied current $j$.
    In (a, b) we fix $\mu H=0.5D^2/J$, and
    in (c, d) we fix $\phi=45^\circ$.
    The velocity color map ranges from dark blue through
    white to dark red, while the annihilation probability
    color map ranges from white to dark red.
    In the hatched regions in (a, c), there are
    no surviving skyrmions present.
  }
  \label{fig:8}
\end{figure}

To summarize our findings,
in Fig.~\ref{fig:8}(a) we plot a heat map of $\langle v_x\rangle$
as a function of $\phi$ versus $j$ for systems
with $\mu H=0.5D^2/J$.
The velocity reaches its greatest magnitude for low
values of $\phi$, but due to
the absence of a hard flow direction for $\phi \leq 10^\circ$,
no diode effect is present.
A hard flow direction emerges once
$\phi>10^\circ$, and the diode effect becomes visible, as
indicated by the appearance of lower velocity magnitudes for
negative currents than for positive currents.
In the small hatched region
with $\phi\leq 30^\circ$ and $j\leq -4\times10^{10}$~A~m$^{-2}$,
all of the skyrmions annihilate; very large current amplitudes
must be applied to achieve this annihilation.
The diode effect occurs across most of the $\phi$ and $\pm j$ pairs
in the phase diagram, but the intensity of the diode motion varies
with $\phi$.
In particular,
the diode effect weakens as $\phi$ approaches $\phi=90^\circ$,
and is lost
at $\phi=90^\circ$ where $\langle v_x\rangle$ becomes symmetric
for positive and negative currents.
We expect that a mirrored version of the heat map would appear for
protrusion defect angles
$\phi>90^\circ$.
The corresponding heat map of the
annihilation rate $p_0$, plotted as a function of
$\phi$ versus $j$ in Fig.~\ref{fig:8}(b),
shows that there is a
higher annihilation rate for negative currents,
indicated by
the darker areas in the $j<0$ region.

In Fig.~\ref{fig:8}(c) we show a heat map of $\langle v_x\rangle$
as a function of $\mu H$ versus $j$ at fixed $\phi=45^\circ$,
and the corresponding heat map of $p_0$ as a function of
$\mu H$ versus $j$ appears in
Fig.~\ref{fig:8}(d).
The diode effect is present
for all values of $\mu H$, as indicated by
Fig.~\ref{fig:8}(c), but the magnitude of the diode effect
increases with
increasing $\mu H$,
as shown by the diminishing magnitude of
$\langle v_x\rangle$ with increasing $\mu H$ in the $j<0$ region
of the plot.
This reduction in the magnitude of the velocity with increasing
magnetic field is a consequence of the stiffening of the skyrmions
and their diminished ability to deform as the magnetic field
becomes higher.
Figure~\ref{fig:8}(d) shows that the annihilation rate
is greater for negative currents, as indicated by the darker areas
in the $j<0$ portion of the plot.
The annihilation rate also increases with increasing magnetic field,
and complete annihilation of all skyrmions occurs
for large current amplitudes in the high magnetic field regime.
This behavior is associated with the brittleness of the skyrmions.
At higher magnetic fields, the skyrmions become smaller and more rigid;
however, under a sufficiently large dragging force produced by the
applied current, they can become brittle and annihilate upon
interacting with the protrusion defects.

\section{Summary}

Using atomistic simulations, we investigated the dynamics of
multiple skyrmions interacting with anisotropy defects
forming a linear protrusion arrangement under adiabatic
spin-transfer-torque
currents.
We observe a diode effect in which the skyrmion velocity $|\langle v_x\rangle|$
is
large for positive $j>0$ currents applied along the easy or $+x$ direction but
reduced for negative $j<0$ currents applied along the hard or $-x$ direction.
The flow direction strongly affects the skyrmion annihilation
rate $p_0$, with annihilation occurring more rapidly and beginning at
lower current magnitudes for negative currents than for positive
currents.
The annihilation rate increases
rapidly with increasing current magnitude for
low $|j|$, and saturates toward $p_0=1$ as $|j|$ increases.
Skyrmions flowing along the easy $+x$ direction experience
a Magnus velocity boost during their entire motion, while
skyrmions flowing along the hard $-x$ direction do not.
The presence of skyrmions pinned in the corners of the protrusion
defects is necessary to facilitate the motion of the skyrmions
in the hard $-x$ direction, since collective interactions
between skyrmions of oscillating size are needed to push the moving skyrmions
over the hard direction barrier.
If the annihilation rate is too large, no hard direction flow can
occur because not enough skyrmions are left in the system, and the
remaining skyrmions are all pinned by the defects.
The value of the protrusion angle $\phi$ strongly affects
$\langle v_x\rangle$ for easy $+x$ flow since a reduction in
$\phi$ increases the skyrmion velocity
due to the Magnus velocity boost.
For negative $-x$ flow,
$\langle v_x\rangle$ is largely insensitive to the $\phi$.
Larger $\phi$ values decrease the skyrmion annihilation
rate for both positive and negative currents,
The exception is vertical $\phi=90^\circ$
protrusions, which have the greatest annihilation rate and also
show no diode effect due to the lack of asymmetry.
When we decrease the magnetic field,
for positive currents
$\langle v_x\rangle$ is slightly reduced, while
for negative currents the window of finite
$\langle v_x\rangle$ values is significantly extended.
Thus, applying a larger magnetic field
enhances the diode effect by
reducing the range of $j$ for which motion along the hard
axis can occur.
Larger magnetic fields stabilize more rigid skyrmions, while
smaller magnetic fields produce softer skyrmions that can deform
more easily and are better able to travel in the $-x$ direction.
The skyrmions also become more brittle when the magnetic field
is large, and the annihilation rate $p_0$ is enhanced.
Under extremely
large magnetic fields, there is complete annihilation of the
skyrmions for large current amplitudes.
We expect our results to be useful for designing
skyrmion diode devices. The velocity dependence
of $\phi$ together with the ability to control the range of $j$
over which hard direction motion occurs
by changing $\mu H$ can provide a device with fine tuning mechanisms.


\section*{Acknowledgments}
This work was supported by the US Department of Energy through the Los Alamos
National Laboratory. Los Alamos National Laboratory is operated by
Triad National Security, LLC, for the National Nuclear Security
Administration of the U. S. Department of Energy (Contract
No. 892333218NCA000001).
J.C.B.S and N.P.V. acknowledge funding from Fundação de Amparo à
Pesquisa do Estado de São Paulo - FAPESP (Grants J.C.B.S 2023/17545-1
and 2022/14053-8, N.P.V 2024/13248-5).
We would like to thank FAPESP for providing the computational
resources used in this work (Grant: 2024/02941-1).
\section*{Data Availability Statement}
Data available on request from the authors.

\section*{References}
\bibliographystyle{iopart-num}
\bibliography{mybib}

\end{document}